# Magnetic Orbital Hall Effect in Altermagnet $RuO_2$

Badsha Sekh[1], Hasibur Rahaman[1], Shilei Ding[1], Pinkesh Kumar Mishra[1], Ramu Maddu[1], Tianli Jin[1], Subhakanta Das[1], and S.N. Piramanayagam[1,*]

[1]School of Physical and Mathematical Sciences, Nanyang Technological University, 21 Nanyang Link, 637371, Singapore

*Corresponding author: prem@ntu.edu.sg*

Orbital angular momentum provides an alternative channel for current-induced magnetization switching beyond conventional spin-orbit coupling. While orbital Hall effects have been observed in several nonmagnetic materials, their manifestation in symmetry-compensated magnetic systems remains unexplored. Here, we report experimental evidence for a magnetic orbital Hall effect in $RuO_2$. In $RuO_2$(101)/Pt/Co heterostructures, we observe a pronounced unconventional torque characterized by a large out-of-plane component, strong crystalline anisotropy, and deterministic field-free switching of a perpendicular ferromagnet, for a large range of $RuO_2$ thickness. The torque exhibits a non-monotonic dependence on Pt thickness with a maximum at 1.5 nm and shows a long-range $RuO_2$ thickness (t) dependence that saturates at $t_{RuO_2} >$ 100 nm. These features cannot be reconciled with conventional spin-current mechanisms. Rather, they indicate a magnetic orbital Hall effect in $RuO_2$ that could originate from exchange-induced momentum-dependent band splitting and its interplay with spin-orbit and crystal-field coupling, with the generated orbital current converted into torque in Pt. Our findings establish altermagnets as intrinsic sources of orbital currents and extend orbitronics to symmetry-compensated magnetic systems.

The generation and manipulation of spin-polarized currents underpin modern spintronic technologies by enabling current-driven torques on adjacent magnets. In conventional spin-orbit torque systems, a longitudinal charge current is converted into a transverse spin current through relativistic spin-orbit coupling (SOC) mechanisms, such as spin Hall effect [1], [2] (SHE) in heavy metals and the spin Rashba-Edelstein effect [3], [4] at inversion-asymmetric interfaces. In these SOC-based platforms, spin degeneracy is lifted, and spin polarization becomes locked to electron momentum, providing efficient pathways to manipulate magnetization for memory [5], [6], [7] and logic devices [8], [9]. Beyond these relativistic mechanisms, the recent discovery of altermagnets [10], [11], [12] has revealed a fundamentally different route to generate spin-polarized currents: momentum-dependent spin splitting emerges from symmetry-compensated antiferromagnetic order, producing nonrelativistic spin-split band structures despite vanishing net magnetization (Fig.1(a)). The resulting spin polarization is tied to the Néel vector, allowing efficient control of spin polarization via magnetic order and crystal symmetry. This enables symmetry-selected spin responses, including out-of-plane components that are essential for field-free deterministic switching of perpendicular magnets.

Electrons in solids carry not only spin but also orbital angular momentum [13]. Although orbital moments are often regarded as quenched by crystal-field effects, recent theoretical [15], [16] and experimental studies [17], [18] have shown that an applied electric field can generate nonequilibrium orbital angular momentum through the orbital character and hybridization of Bloch states in crystalline solids. Unlike the spin Hall effect, the orbital Hall response originates from momentum-space orbital texture [19] and does not rely on relativistic spin-orbit coupling. As a result, the orbital Hall conductivity can be remarkably large in light materials [20], [21], often exceeding conventional spin Hall conductivities [22], [23]. Moreover, orbital angular momentum can propagate over long distances and, after conversion into spin polarization inside a ferromagnet (FM), can exert torques beyond the typical spin dephasing length. This long-range orbital transport opens opportunities for manipulating and switching thermally stable, thicker magnetic layers that are challenging to control using conventional spin-orbit torques.

Although the nonrelativistic spin splitting [24], [25] in altermagnets and the orbital Hall effect [26], [27] in nonmagnetic materials have been extensively studied in the past few years, their interplay remains largely unexplored. In nonmagnetic systems, time-reversal symmetry restricts the orbital Hall response by constraining the allowed components of the orbital conductivity. The onset of magnetic order breaks time-reversal symmetry, lowers the system symmetry, and enables additional orbital response components that are not permitted in nonmagnetic materials. To date, such magnetic orbital Hall effects have been investigated mainly in simple transition ferromagnets [21], [28], leaving unresolved how spin splitting compensated magnets, such as altermagnets, modify orbital transport and the resulting torque generation.

In this Letter, we demonstrate a magnetic orbital Hall effect in altermagnetic $RuO_2$(101)/Pt/Co heterostructures. We observe a pronounced out-of-plane torque, deterministic field-free switching and a strong anisotropy governed by the in-plane current direction. Critically, the torque magnitude shows a non-monotonic dependence of Pt thickness and a long range RuO2 thickness dependence. The torque reaches a maximum value of 3.68× $10^5$ $\Omega^{-1}m^{-1}$ at an intermediate Pt thickness of 1.5 nm, consistent with optimized orbital-to-spin conversion in the Pt interlayer, and exhibits clear crystalline anisotropy, with the largest response for current applied along the [010] crystal direction. Moreover, the out-of-plane torque efficiency increases with $RuO_2$ thickness and saturates above 100 nm, indicating a long diffusion length characteristic of orbital transport. These results provide the first experimental demonstration of a magnetic orbital Hall effect in an altermagnet, positioning altermagnets as a new symmetry-driven platform for engineering anisotropic orbital currents.

To investigate the magnetic orbital Hall effect in $RuO_2$, epitaxial $RuO_2$(101) thin films were deposited on single-crystal $Al_2O_3(1\bar{1}02)$ substrates at a substrate temperature of 450°C. X-ray diffraction and reciprocal space mapping confirm the high epitaxial quality of the $RuO_2$ layer (see Fig.S1 in Supplemental Material). Subsequently, Pt ($t_{Pt}$)/Co (1 nm)/Ru (2 nm) layers were sputtered at room temperature on the $RuO_2$ films. The Ru(2 nm) layer purely serves as a capping layer to prevent oxidation of the Co layer. It is to be noted that there is no heavy-metal layer between Co and Ru, and hence any potential orbital contribution from this thin Ru layer could not cause any significant contribution to the magnetization switching results. Magnetic characterization reveals perpendicular magnetic anisotropy in the $RuO_2$/Pt/Co stacks when the Pt thickness is in the range of 1.2-2.0 nm (see Fig. S2 in the Supplemental Material).

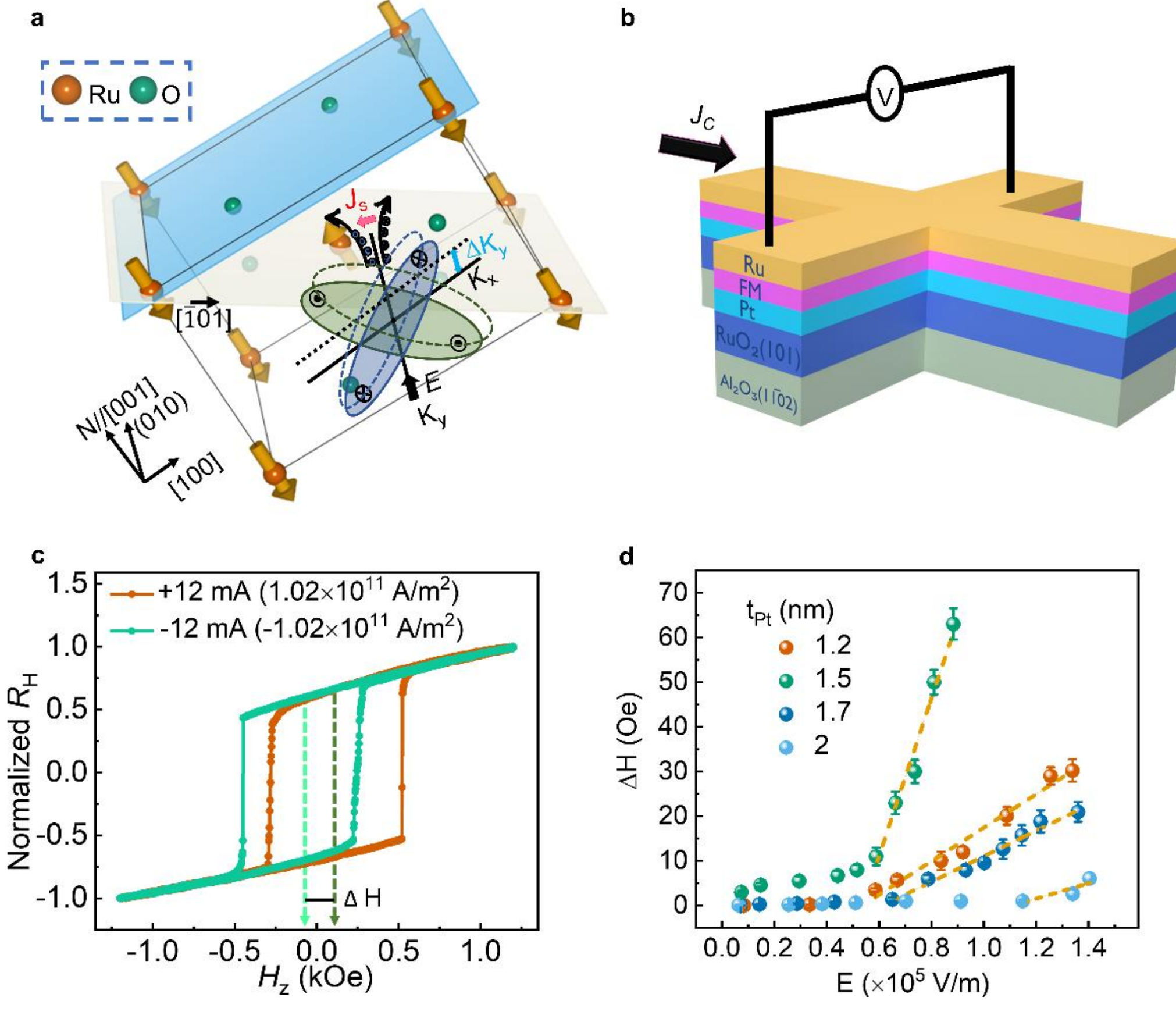

Fig. 1. (a) Crystal and magnetic structure of $RuO_2$, with the Ru magnetic moments aligned along the [001] direction. Momentum-dependent spin splitting arises from the broken $T\tau$ symmetry in the (001) plane. An applied electric field along 90° crystal angle with respect to the $[\bar{1}01]$ direction generates a transverse spin current with a finite projection along [001], producing an out-of-plane torque. (b) Schematic of the Hall bar device used for current-induced torque and switching measurements. (c) Normalized anomalous Hall resistance ($R_H$) as a function of out-of-plane magnetic field ($H_z$) under DC electric field of $0.88\times10^5$ V/m (±12 mA current) at zero in-plane field for the $RuO_2$/Pt (1.5)/Co (1)stack. The dotted lines indicate the shift of the hysteresis loop centers, and the separation between them ($\Delta$H) quantifies the effective out-of-plane magnetic field. (d) Dependence of $\Delta$H on applied electric field (E) for samples having different Pt thicknesses. The dotted lines are linear fits at higher E values.

The films were then patterned into Hall bar devices with a length of 30 μm and a width of 5 μm using photolithography and Ar ion etching for transport measurements. All measurements were performed at room temperature. Transport measurement schematics and the layer structure are shown in Fig. 1(b). A DC current is applied along the Hall bar to probe the corresponding anomalous Hall response. We first verify the presence of an unconventional current-induced torque arising from z-polarized angular momentum. In the presence of a z-polarized spin or orbital current, the generated out-of-plane damping-like torque competes with the intrinsic damping of the perpendicular magnet. When the applied current is sufficiently large, this damping-like torque produces a measurable shift of the anomalous Hall hysteresis loop [29], [30], [31], analogous to an effective exchange-bias field[32]. By applying positive and negative electric fields of $0.88\times10^5$ V/m (corresponding to ±12 mA) in $RuO_2$/Pt/Co, we observe a clear hysteresis loop shift defined as $\Delta$H = $H_{center}$ (+I) $H_{center}$ (−I), where $H_{center}$ = ($H_r$+ $H_l$)/2, and $H_r$ and $H_l$denote the right and left coercive fields, respectively (Fig. 1(c)).

Fig. 1(d) shows the electric field dependence of $\Delta$H for $RuO_2$/Pt(1.2, 1.5, 1.7, 2 nm)/Co samples. Above a threshold current, $\Delta$H varies linearly with the applied electric field. Notably, this loop shift occurs in the absence of any external in-plane magnetic field, in contrast to conventional spin-orbit torque mechanisms that require in-plane symmetry breaking to generate an effective damping-like torque. The effective damping-like torque efficiency per unit electric field [33], [34] (E) is extracted from the slope of $\Delta$H/E using $\xi_{DL}^{E} = \frac{2e}{\hbar} \mu_0 M_s t_{FM} \frac{\Delta H}{E}$ , where $M_s$ and $t_{FM}$ are the saturation magnetization and thickness of the Co layer, respectively. The extracted torque efficiency reaches $3.68\times 10^5\ \Omega^{-1}m^{-1}$ for $RuO_2$/Pt(1.5)/Co.

The observed torque efficiency is nearly an order of magnitude higher than prior experimental reports [10], [35] on the spin splitting torque in $RuO_2$. Such a strong unconventional torque cannot be simply attributed to spin-splitting torque originating from $RuO_2$. The insertion of a Pt interlayer indeed introduces a conventional spin Hall channel. However, in our symmetric Hall-bar geometry and layer configuration, the spin Hall effect in Pt cannot generate a net out-of-plane spin current without additional in-plane symmetry breaking [36]. Furthermore, if the torque were dominated by spin current transport arising from spin splitting in $RuO_2$, the Pt interlayer would be expected to reduce rather than enhance the signal. The spin diffusion length in Pt is typically[37] on the order of 1-2 nm. In this case, increasing the Pt thickness should progressively reduce the spin current transmitted from $RuO_2$ to the Co layer, leading to a monotonic decrease of torque efficiency. Even if one neglects spin diffusion and considers only the additional y-polarized spin current generated within Pt via the spin Hall effect, its contribution would result in at most a modest (~10%) enhancement of the overall torque efficiency (Fig.S4). This small enhancement cannot explain the pronounced peak in torque efficiency observed at 1.5 nm Pt. In contrast, our measurements reveal a clear non-monotonic dependence of torque efficiency on Pt thickness, with a pronounced maximum at 1.5 nm. Moreover, a Pt thickness exceeding 1 nm is sufficient to suppress direct exchange coupling between $RuO_2$ and Co, thereby ruling out interfacial exchange-driven[38] mechanisms as the origin of the unconventional torque.

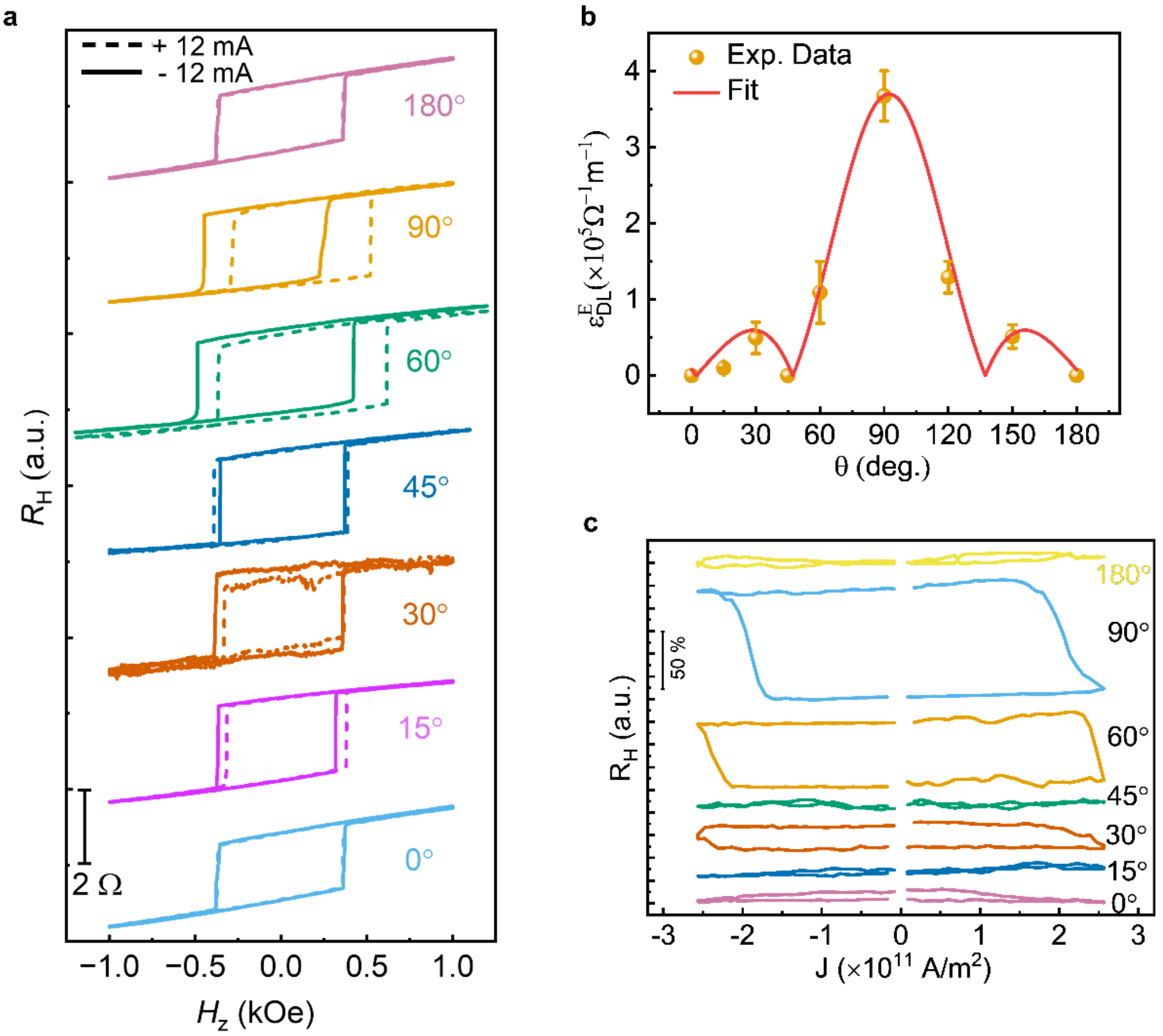

Fig. 2. (a) Normalized anomalous Hall resistance ($R_H$) as a function of out-of-plane magnetic field ($H_Z$) (I= ±12 mA and at zero in-plane field) for different in-plane current directions. The charge current (or electric field) $J_c$ is applied at crystal angles θ with respect to the [$\bar{1}$01] axis, where $\theta$ = 0°,15°,30°,45°,60°,90°,180°. (b) Angular dependence of the unconventional torque efficiency extracted from the hysteresis loop shift measurements as a function of θ. (c) Current-induced switching loops measured as $R_H$ versus applied current density (J) without any external in-plane magnetic field for devices with different θ. The applied current density ranges from $-2.56 \times 10^{11}$ to $+2.56 \times 10^{11}$ A/m$^2$.

The observed thickness dependence instead follows the characteristic behavior of orbital-current generation in $RuO_2$ combined with orbital-to-spin conversion in Pt. Orbital angular momentum generated in $RuO_2$ can propagate across the interface and be converted into spin polarization within Pt through spin-orbit coupling. An optimal Pt thickness of approximately 1.5 nm maximizes this orbital-to-spin conversion efficiency, while thicker Pt layers introduce additional momentum and spin relaxation processes that reduce the transmission of the converted angular momentum to the Co layer. Taking together, these observations strongly support an orbital current-mediated torque mechanism rather than a purely spin current origin.

To further elucidate the origin of the unconventional torque, we investigate its dependence on the in-plane current direction. Multiple Hall-bar devices were fabricated with the longitudinal axis oriented at different crystallographic angles θ relative to [$\bar{1}$01] direction. By applying a constant DC current of ±12 mA, we measured the corresponding hysteresis loop shifts for each orientation, as shown in Fig. 2(a). A pronounced angular dependence is observed, where the torque vanishes at θ = 0°, corresponding to current applied along the [$\bar{1}$01] direction, while it reaches a maximum at θ = 90°, corresponding to current along the [010] axis. This strong anisotropy is consistent with angular-momentum generation governed by the momentum-dependent spin splitting of the altermagnetic $RuO_2$ band structure. In altermagnets, spin polarization is locked to crystal momentum through symmetry-protected spin splitting. When a charge current is applied along the [010] direction, the nonequilibrium carrier distribution shifts in momentum space, producing a net spin polarization through spin-momentum locking. In this configuration, the resulting transverse spin current flows predominantly along the [100] direction. Because the $RuO_2$ (101) film is oriented such that the [100] axis contains a component perpendicular to the film plane, the generated spin current acquires an out-of-plane flow component that can be injected into the adjacent Co layer. The associated spin polarization is oriented along the [001] axis, giving rise to an out-of-plane damping-like torque.

In contrast, when the current is applied along the [$\bar{1}$01] direction (θ = 0°), symmetry prevents the generation of a transverse spin current with a component normal to the $RuO_2$(101) plane. Although a projection of the current within the (001) plane can still induce spin polarization through momentum-dependent splitting, the resulting transverse spin current flows parallel to the film plane (along [010]), lacking any component perpendicular to the interface. As a result, no angular momentum is transferred into the Co layer, and no measurable torque is observed.

Fig. 2(b) presents the angular dependence of the effective torque efficiency. A pronounced anisotropy is observed: the torque exhibits a minimum at θ = 0° and near 45°, reaches a maximum at 90°, and shows a secondary maximum around 30°. This nontrivial angular profile directly reflects the symmetry-filtered generation of angular momentum in altermagnetic $RuO_2$, where both the polarization direction and the propagation direction of the generated transverse spin current are determined by the crystallographic orientation of the applied current.

To quantitatively describe this behavior, we model the angular dependence by considering the current-angle dependent out-of-plane spin current component normal to the $RuO_2$(101) plane. The torque efficiency ($\tau(\theta)$) follows as

$$\tau(\theta) \propto |\sin\theta \ \frac{cos^2\theta-2sin^2\theta}{cos^2\theta+2sin^2\theta}| \qquad (1)$$

which captures the symmetry-imposed projection of the transverse spin current onto the film's normal direction. The solid curve in Fig. 2(b) shows the fitting based on the equation. (1), demonstrating excellent agreement with experimental data.

The angular dependence of the torque efficiency is further validated by field-free current-induced switching measurements shown in Fig. 2(c). Efficient magnetization reversal of the Co layer is achieved when the current is applied along the [010] direction, with a switching ratio of approximately 90 %, whereas switching is strongly suppressed for current orientations near 0°. These results consistently confirm that the unconventional torque originates from symmetry-selected angular-momentum generation in the altermagnetic $RuO_2$ layer.

In order to probe the contributions from orbital torque in $RuO_2$ altermagnets, we made a series of samples with different $RuO_2$ thickness. The XRD peak intensities arising from epitaxially grown $RuO_2$ (101) crystal planes increase with increasing $t_{RuO_2}$, (Fig. 3(a)), indicating enhanced scattering volume with thickness. To compensate for the altered interfacial and FM layer anisotropy conditions at larger $RuO_2$, $[Co(0.6)/Pt(0.8)]_{\times 2}$ was used as the ferromagnetic layer to preserve sharp PMA (Fig. 3(b)). Fig. 3(c) presents the unconventional torque efficiency in $RuO_2$/Pt/FM as a function of $RuO_2$ thickness. The torque efficiency (ξ) increases monotonically with increasing $RuO_2$ thickness and gradually saturates when the thickness exceeds ~100 nm. Within a simple drift-diffusion framework [39], the thickness dependence of the torque can be described by the equation [40],

$$\Xi(t)=\xi_{max}\,[1-\mathrm{sech}(t/\lambda)] \qquad (2)$$

where $\xi_{max}$ is the maximum SOT switching efficiency (see details in S4 of the Supplemental Materials) and λ is the effective diffusion length of the transported angular momentum. Fitting the experimental data yields λ =36.8 ± 2 nm. This value is substantially larger than the reported [35], [41] spin diffusion length in $RuO_2$, but is consistent with the long propagation length previously associated with orbital currents. The extracted diffusion length, therefore, strongly supports the interpretation that the unconventional torque originates from the transport of z-polarized orbital current [42] rather than conventional spin current.

Furthermore, the enhancement of torque efficiency with increasing $RuO_2$ thickness is also proved by field-free switching measurements (see the Supplemental Materials for details). Devices with thicker $RuO_2$ layers exhibit significantly reduced switching power consumption, reflecting the improved torque efficiency (Fig. 3(d)). This thickness-dependent scaling of both torque efficiency and switching power provides additional evidence for an orbital transport mechanism underlying the unconventional torque.

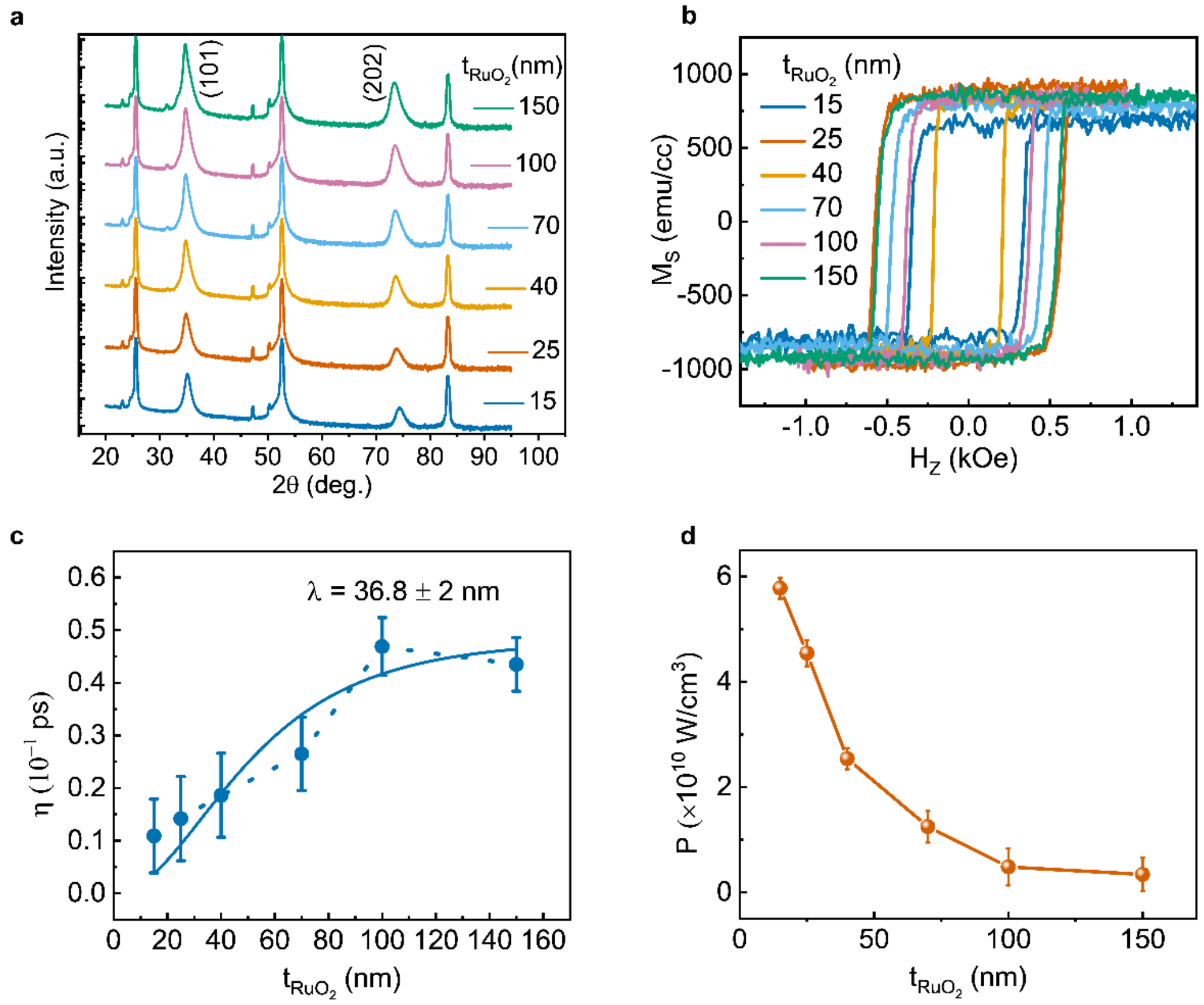

Fig. 3 (a) Gonio scan (θ-2θ) showing the crystallinity of the sputtered $RuO_2$ films of various thicknesses (($t_{RuO_2}$). (b) Out of plane (OOP) magnetic hysteresis loops ($t_{RuO_2}$) show sharp PMA across large $t_{RuO_2}$range. (c) $RuO_2$ thickness dependence of the unconventional torque efficiency in $RuO_2$/Pt/FM, exhibiting saturation at large thickness consistent with long diffusion length. The solid curve is a fit using equation (2). (d) The enhanced torque efficiency leads to a substantial reduction in switching power.

As discussed above, a spin current generated solely from nonrelativistic spin splitting in $RuO_2$ cannot account for either the magnitude or the long diffusion length of the observed torque in $RuO_2$/Pt/FM, although the angular dependence of the torque follows the symmetry of the spin-splitting mechanism. A purely spin-based picture would produce a spin current dictated by the momentum-dependent spin texture, but it does not naturally explain the optimization with Pt thickness. These observations indicate that an additional channel of angular-momentum transport must be involved.

The emergence of orbital current in $RuO_2$ can be understood from the microscopic electronic structure of altermagnets. In equilibrium, orbital angular momentum (OAM) in transition-metals is largely quenched by the crystal field. However, as emphasized in the orbitronics framework, OAM is excited under an external electric field through interband coherence [43] and redistribution of Bloch states. In $RuO_2$, the states near the Fermi level are dominated by Ru 4d t2g orbitals subject to an orthorhombic (D2h) crystal field [44]. The exchange-driven, momentum-dependent spin splitting characteristic of altermagnetism reshapes the orbital composition of these Bloch states.

As a result, the spin-split bands acquire momentum-dependent orbital expectation values, producing a nontrivial orbital texture in k-space.

This mechanism has been explicitly analyzed for $RuO_2$ by Yahagi [45] through theoretical studies without experimental confirmation. The orbital Hall conductivity arises predominantly from the time-reversal-odd term associated with the single-band orbital expectation value, and its magnitude scales with the spin-split electronic structure. Microscopically, spin-orbit interaction mediates the transfer of angular momentum from the exchange-generated spin current[46] to the orbital channel, while the D2h crystal field anisotropically lifts orbital degeneracies and activates the quenched orbital angular momentum, thereby defining the symmetry and strength of the orbital current. As a result, a sizable magnetic orbital Hall response emerges whose symmetry is locked to the Néel vector. When an electric field shifts the Fermi surface, this momentum-dependent orbital texture becomes imbalanced between +k and −k states, generating a net transverse orbital current. In this sense, orbital current generation in $RuO_2$ is a direct consequence of its symmetry-compensated magnetic order combined with its multi-orbital electronic structure.

Complementary to this picture, the spin-orbital altermagnetism [47] framework demonstrates that once orbital degrees of freedom are included, altermagnets generically host momentum-dependent spin-orbital textures even without relying on strong relativistic spin-orbit coupling. In this sense, orbital splitting in the conventional band is not required. Although orbital angular momentum is not a conserved quantum number in a crystal, magnetic order can induce a momentum-dependent orbital polarization through spin-orbital locking [48]. An applied electric field then converts this orbital texture into a transverse orbital current via Fermi-surface shift. Thus, orbital current generation in $RuO_2$ is a natural consequence of its symmetry-compensated magnetic order combined with its multi-orbital electronic structure.

Furthermore, when spin-group symmetry is further reduced in the presence of spin-orbit coupling, additional orbital response components are shown to become symmetry-allowed, including out-of-plane polarized orbital currents, as predicted in recent theory [49]. In our heterostructure, the orbital current generated in $RuO_2$ preserves the angular symmetry imposed by the spin-splitting mechanism, explaining the observed crystalline anisotropy. The Pt interlayer then serves primarily as an efficient orbital-to-spin conversion medium, giving rise to the large out-of-plane torque detected experimentally. In this framework, the unconventional torque in $RuO_2$ based devices is not a simple extension of spin splitting, but rather a manifestation of symmetry-governed magnetic orbital Hall[50] transport intrinsic to altermagnets.

We note that the theoretical analysis by Yahagi predicts a strong dependence of the magnetic orbital Hall conductivity on the Néel vector orientation, with a pronounced response for Néel alignment along [100] and a substantial suppression when aligned along [001] due to the anisotropic activation of orbital angular momentum under the orthorhombic crystal field. In contrast, our experiments reveal a sizable unconventional torque even though the Néel vector in $RuO_2$ is predominantly oriented along [001]. Rather than contradicting the theoretical framework, our results suggest that additional symmetry-allowed orbital transport channels may become operative in realistic heterostructures. In particular, exchange-driven momentum-dependent spin-orbital textures, epitaxial strain in the deposited $RuO_2$ and reduced symmetry at the $RuO_2$/Pt interface could modify the allowed orbital response [51] compared with the ideal bulk configuration.

In summary, we report a pronounced unconventional torque in altermagnetic $RuO_2$ based heterostructures, characterized by a large out-of-plane angular-momentum component, strong crystalline anisotropy, and deterministic field-free switching. The torque exhibits a non-monotonic dependence on Pt thickness with a maximum at 1.5 nm, together with a long-range $RuO_2$ thickness dependence that saturates beyond 100 nm. These behaviors cannot be reconciled with conventional spin-current mechanisms alone, neither the symmetry of the spin Hall effect from Pt nor the short spin diffusion length can account for the large magnitude and long diffusion observed. Instead, the results point to an orbital transport origin, where exchange-driven momentum-dependent band splitting in $RuO_2$ could be the origin of the magnetic orbital Hall current through its interplay with spin-orbit coupling. The orbital current is further converted into spin and delivered as a torque to the magnetization of the adjacent ferromagnet. Our results therefore provide experimental evidence for a magnetic orbital Hall effect in an altermagnet and establish $RuO_2$ as a platform for engineering anisotropic orbital currents and efficient, field-free magnetization switching, opening new opportunities for low-power orbitronic devices.

The authors gratefully acknowledge the funding from the National Research Foundation (NRF), Singapore, for the CRP21 and CRP-Frontier grants (NRF-CRP21-2018-0003, NRF-F-CRP-2024-0012) and Ministry of Education, Singapore, for the Tier 2 Grant MOET2EP50122-0023. BS thanks the NRF and NTU research scholarship for carrying out research at NTU. SD acknowledges the Start-Up Grant from Nanyang Technological University.